\documentclass{article}

\usepackage{microtype}
\usepackage{graphicx}
\usepackage{booktabs} 

\usepackage{hyperref}

\usepackage[accepted]{icml2024}

\usepackage{amsmath}
\usepackage{amssymb}
\usepackage{mathtools}
\usepackage{amsthm}

\usepackage[capitalize,noabbrev]{cleveref}

\theoremstyle{plain}

\theoremstyle{definition}

\theoremstyle{remark}

\usepackage[textsize=tiny]{todonotes}

\icmltitlerunning{Global Gravity Wave Simulation Using ML}

\begin{document}

\twocolumn[
\icmltitle{Machine Learning Global Simulation of Nonlocal Gravity Wave Propagation}



\icmlsetsymbol{equal}{*}

\begin{icmlauthorlist}
\icmlauthor{Aman Gupta}{equal,1}
\icmlauthor{Aditi Sheshadri}{equal,1}
\icmlauthor{Sujit Roy}{equal,2}
\icmlauthor{Vishal Gaur}{2}
\icmlauthor{Manil Maskey}{3}
\icmlauthor{Rahul Ramachandran}{3}
\end{icmlauthorlist}

\icmlaffiliation{1}{Department of Earth System Science, Stanford University, Stanford, USA}
\icmlaffiliation{2}{Earth System Science Center, The University of Alabama in Huntsville, Huntsville, AL, USA}
\icmlaffiliation{3} {NASA Marshall Space Flight Center, Huntsville, AL, USA}


\icmlcorrespondingauthor{Aman Gupta}{ag4680@stanford.edu}

\icmlkeywords{earth system modeling, climate model parameterization, atmospheric gravity waves, machine learning, subgrid scale modeling, atmospheric dynamics}

\vskip 0.3in
]



\printAffiliationsAndNotice{\icmlEqualContribution} 

\begin{abstract}
Global climate models typically operate at a grid resolution of hundreds of kilometers and fail to resolve atmospheric mesoscale processes, e.g., clouds, precipitation, and gravity waves (GWs). Model representation of these processes and their sources is essential to the global circulation and planetary energy budget, but subgrid scale contributions from these processes are often only approximately represented in models using \emph{parameterizations}. These parameterizations are subject to approximations and idealizations, which limit their capability and accuracy. The most drastic of these approximations is the ``single-column approximation" which completely neglects the horizontal evolution of these processes, resulting in key biases in current climate models. With a focus on atmospheric GWs, we present the first-ever global simulation of atmospheric GW fluxes using machine learning (ML) models trained on the WINDSET dataset to emulate global GW emulation in the atmosphere, as an alternative to traditional single-column parameterizations. Using an Attention U-Net-based architecture trained on globally resolved GW momentum fluxes, we illustrate the importance and effectiveness of global nonlocality, when simulating GWs using data-driven schemes. 

\end{abstract}

\section{Introduction}
\label{submission}

Gravity waves are fast-propagating perturbations in a stably stratified fluid. In the atmosphere, they are constantly generated by a myriad of sources like jet imbalance, geostrophic adjustment processes, flow over mountains, storm tracks, etc. Their spatial scales range from O(100) m to O(1000) km, i.e., they span across the atmospheric mesoscales and submesoscales.

\begin{figure*}
\vskip 0.2in
\begin{center}
\includegraphics[width=1.8\columnwidth]{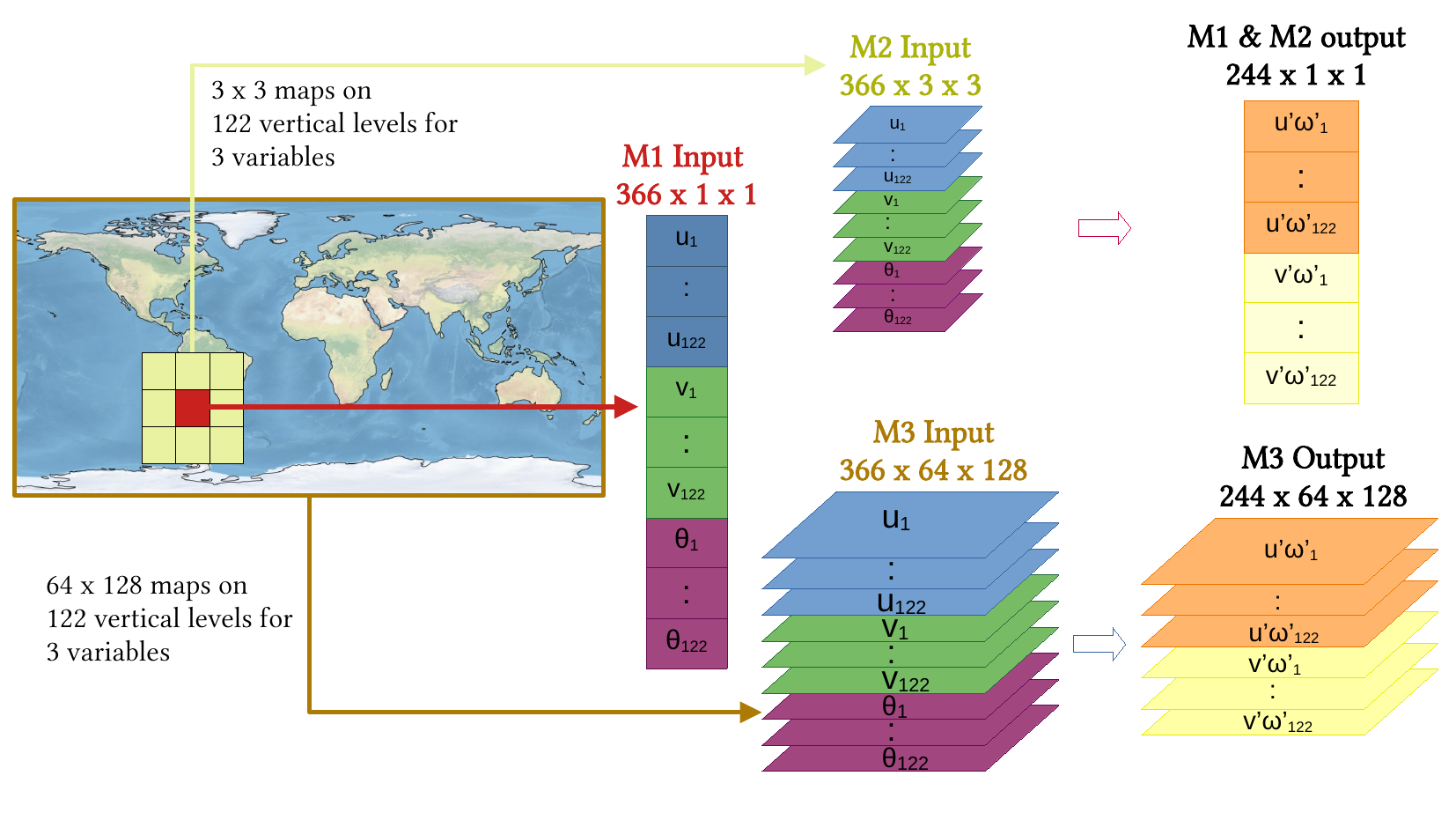}
\caption{The three architectures used for global GW resolved momentum flux simulation. The three architectures, described in section \ref{subsec:model_arch} employ three different degrees of nonlocality. On one end, M1 uses single-column background data to predict the fluxes within that column. A timeslice is therefore a single vector of length 366. Intermediately, M2 uses background information in a 3$\times$3 stencil to predict the fluxes within the single-column at the center of the stencil. A timeslice for M2 has dimensions 3 $\times$ 3 $\times$ 366. On the other end, M3 uses global maps of the background field to predict global maps of fluxes. A timeslice for M3, thus, has dimensions 366 $\times$ 64 $\times$ 128.}
\label{fig:schematic}
\end{center}
\vskip -0.2in
\end{figure*}

Gravity waves (GWs) dynamically couple the different layers of the atmosphere and are among the key drivers of the meridional overturning circulation in the middle atmosphere \cite{Fritts.Alexander2003, Achatz.etal2023}. They are primary contributors in driving the pole-to-pole mesospheric circulation \cite{Holton1982, Becker2012}. In the stratosphere, they influence the quasi-biennial oscillation (QBO) of tropical winds \cite{Giorgetta.etal2002}, and the springtime breakdown of the Antarctic polar vortex \cite{Gupta.etal2021a}. GWs can also contribute to rapid breakdowns of the wintertime polar vortex, i.e., sudden warmings \cite{Albers.Birner2014, Song.etal2020}, eventually influencing tropospheric storm tracks \cite{Kidston.etal2015, Domeisen.Butler2020}.

Due to limited grid resolution, all state-of-the-art climate models represent subgrid momentum fluxes due to GWs using \emph{parameterizations}. Depending on the source, these parameterizations can be broadly classified as orographic (for GWs generated over mountains, having zero ground-based phase speed) and 
nonorographic (generate elsewhere, having non-zero phase speeds). The most prominent orographic parameterizations include \citet{Lott.Miller1997, vanNiekerk.etal2020} and the most prominent nonorographic parameterizations include \citet{Alexander.Dunkerton1999, Scinocca2002, Scinocca2003}.
All these schemes use the large-scale background state resolved by the climate models to predict the subgrid-scale momentum fluxes. The generated momentum fluxes are then coupled with the large-scale momentum equations that solve for the resolved flow dynamics in the model. Over nearly four decades now, all parameterizations have employed the single-column approximation, i.e., only the atmospheric state within a model vertical column is used to determine the GW flux in that column, thus neglecting any horizontal propagation that these waves exhibit. This assumption directly contradicts observations \citep{Sato.etal2009, Sato.etal2012, Geldenhuys.etal2023} and mesoscale resolving simulations \cite{Kruse.etal2022, Hindley.etal2020, Gupta.etal2024}, that show that GWs can often propagate horizontally thousands of kilometers away from their sources. Past studies have often reiterated the limitations of these assumptions and the urgent need to represent lateral propagation to resolve key circulation biases resulting from these assumptions \cite{McLandress.etal2012,Cámara.etal2016, Kruse.etal2022, Kim.etal2024, Gupta.etal2024a}.

Although WKB ray-tracing-based \cite{Amemiya.Sato2016, Voelker.etal2023} and momentum redistribution-based schemes \cite{Eichinger.etal2023} provide viable alternatives to represent lateral propagation by simulating wave trajectories along which they conserve pseudomomentum, these schemes continue to face computational roadblocks.

Machine learning provides a promising, computationally efficient avenue to generate a new class of data-driven PDE solvers and model parameterizations that learn both large-scale and subgrid-scale physics, directly from high-quality data \cite{Mansfield.etal2023, Roy.etal2024}. Such ML schemes can be trained to take the background atmospheric state as input (just like traditional parameterizations) and use the state to predict the subgrid-scale momentum fluxes. These ML models can subsequently be coupled with the traditional Fortran-based momentum equations solvers. This effectively transforms the problem from the parameterization tuning and approximate modeling space into a problem that focuses on the development of physics-informed ML architectures and their optimal training on high-fidelity data.

This study focuses on the development of such parameterizations. Unlike existing models which have been trained on GW parameterization output, the goal here is to develop ML simulators that learn from inter-annual records of \emph{resolved} momentum fluxes derived from modern reanalysis and kilometer-scale global climate models. The first step involves the training of ML models followed by offline testing of the inferred fluxes. The following step involves coupling these data-driven predictors to coarse-resolution models to test their online performance. Here, we focus on the first step.

The data-driven GW scheme discussed here is also being prepared to be used as part of a much larger foundation model for weather and climate, where it serves as a downstream application to first predict the global atmospheric state and then infer the small-scale GW flux distribution corresponding to that state.

\begin{figure*}
\vskip 0.2in
\begin{center}
\includegraphics[width=1.9\columnwidth]{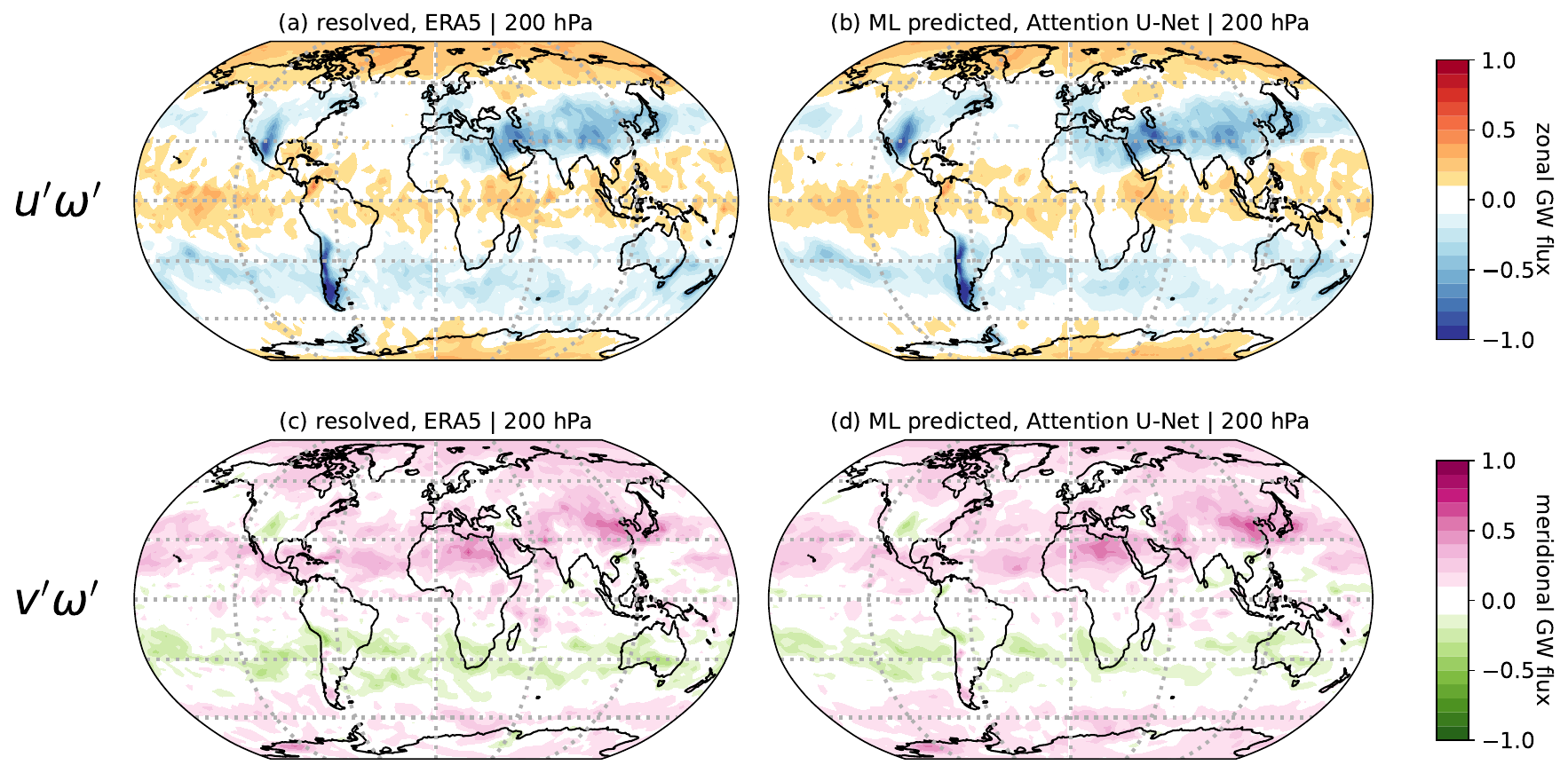}
\caption{Mean predicted fluxes from globally nonlocal model, M3, for May 2015 at 200 hPa height. (a) and (b) respectively show the true mean and the predicted mean zonal flux ($u'\omega'$) for May 2015. (c) and (d) show the true mean and the predicted mean meridional flux ($v'\omega'$). The WINDSET dataset contains input variables and momentum fluxes which were normalized using a constant mean and standard deviation. Mean predicted fluxes for Models M1 and M3 are shown in Figures \ref{fig:comp_zonal} and \ref{fig:comp_meridional}.}
\label{fig:flux_comparison}
\end{center}
\vskip -0.2in
\end{figure*}

\subsection{Previous Works}

ML emulation of gravity wave forcing in climate models has been explored in the past \cite{Chantry.etal2021, Espinosa.etal2022, Lu.etal2024, Sun.etal2023}. However, such efforts have focused on learning GW fluxes from parameterized drag, not resolved drag. Parameterized fluxes have highly analytical forms and contain biases. As a result, the ML models trained on such parameterizations, while easier to train, themselves contain these biases; pure-vertical GW propagation being the most prominent bias.

\citet{Wang.etal2022} proposed a strategy to embed nonlocality within ANNs by using input data from surrounding columns to infer the momentum fluxes within a given column. The strategy can be useful when learning from resolved GW momentum fluxes, as opposed to single-column output. This study, in part, builds upon their idea and explores different degrees of embedded spatial nonlocality: single-column, neighboring cells, and global nonlocality.

\section{Methodology}
\subsection{Dataset}
The training uses the ``Weather Insights and Novel Data for Systematic Evaluation and Testing" (WINDSET) data introduced by \citet{Shinde.etal2024}. WINDSET is a compilation of multiple weather-related datasets incl. long-term precipitation forecasting, hurricane prediction and intensity estimation, aviation turbulence prediction, natural language forecasting, etc. It also comprises four years of the background field and resolved GW momentum fluxes derived from modern reanalysis, ERA5 \cite{Hersbach.etal2020}.

The GW momentum fluxes in WINDSET used ERA5 at its native 30 km resolution to compute the background atmospheric state, and GW fluxes using Helmholtz decomposition, and conservatively coarsegrained the input fields and output fluxes to a 64 $\times$ 128 
(latitude x longitude) Gaussian grid and 137 model levels. The 15 vertical levels near the model top are removed to eliminate artificial damping effects, and thus there are 122 vertical levels.

The input comprises the meteorological variables: zonal wind ($u$), meridional wind ($v$), potential temperature ($\theta$). The vertical velocity ($\omega$) is not added because the hydrostatic model allows only two degrees of freedom. $\theta$ serves as an appropriate vertical coordinate that combines both temperature and pressure information. The output comprises the zonal and meridional components of the vertical momentum flux ($u'\omega'$ and $v'\omega'$). The variables are stacked along the vertical dimension. Thus, an input timeslice has dimensions. 366 $\times$ 64 $\times$ 128 and an output timeslice has dimensions 244 $\times$ 64 $\times$ 128.

The data is available for four years: 2010, 2012, 2014, and 2015 at an hourly resolution. For single-column ANN training, this corresponds to $\approx$300 million data samples for training (not spatio-temporally uncorrelated). For global training, this corresponds to roughly 35k training samples.

\subsection{Model Architecture}
\label{subsec:model_arch}
When dealing with nonlocal propagation of mesoscale systems, one pressing question arises naturally: \emph{how much spatial nonlocality should the ML model represent?}. To address this, we train a set of three ML models which consider different degrees of nonlocality in their input:
\begin{enumerate}
    \item[M1.] Single Column (1 $\times$ 1) ANN: with 4 hidden layers, each twice the input layer size (366), using ReLU activation, Adam optimized with cyclic learning rates. This single-column model aims to replicate the design for traditional single-column parameterizations.
    \item[M2.] A nonlocal (3 $\times$ 3) ANN-CNN: that predicts the fluxes in a given column using the 3 $\times$ 3 grid surrounding the column. The first (input) layer is a 3 $\times$ 3 convolution layer which pools the data into a single column.
    \item[M3.] Global Attention U-Net utilizing convolution layers \cite{Oktay.etal2018}: that takes global (64 $\times$ 128) data with 366 input channels as input, encodes it using a U-Net backbone with residual connections scaled with attention multipliers, and decodes it to produce global flux predictions with 244 output channels. 
\end{enumerate}

\begin{figure}[!ht]
\vskip 0.2in
\begin{center}
\includegraphics[width=0.9\columnwidth]{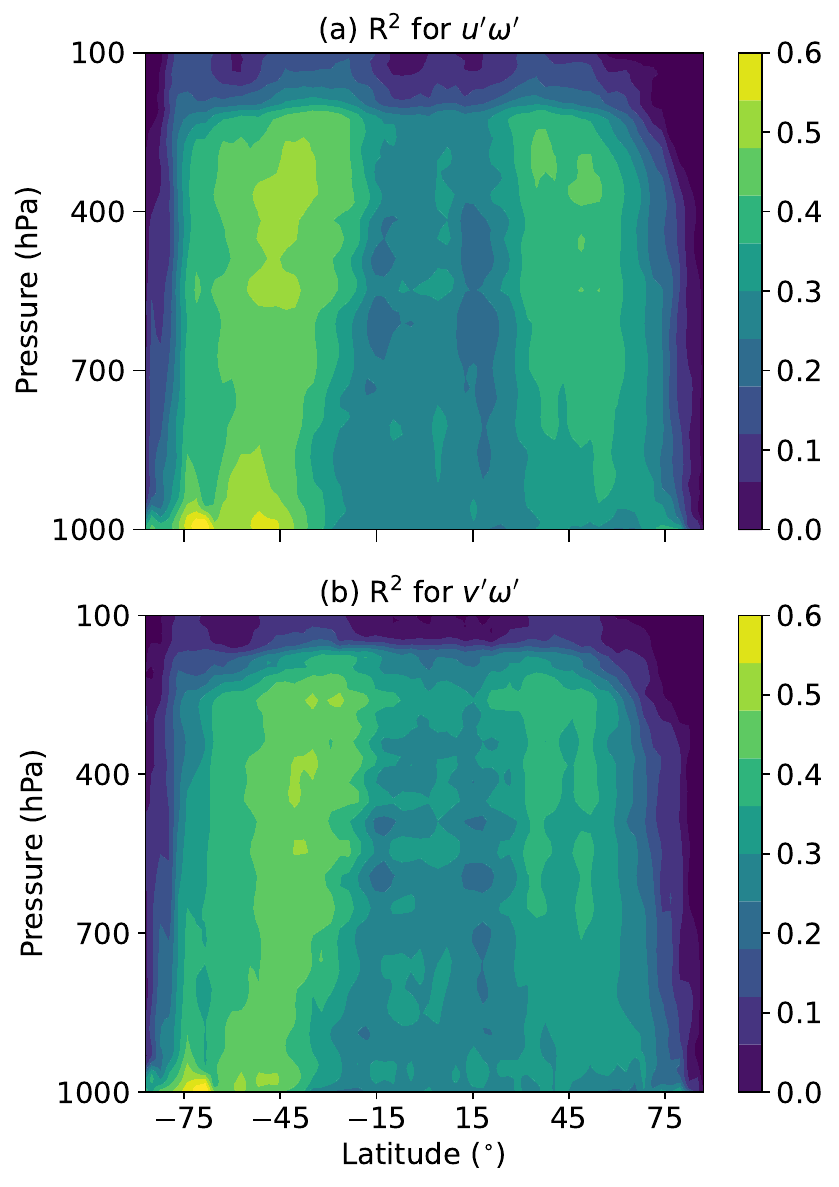}
\caption{R$^2$ value for M3 for (a) zonal flux and (b) meridional flux predictions for May 2015. R$^2$ denotes the percent variance captured by the predictor. A higher R$^2$ value indicates better prediction. }
\label{fig:r2}
\end{center}
\vskip -0.2in
\end{figure}

The architectures are illustrated in Figure \ref{fig:schematic}. The models were optimized to minimize the mean squared error. The models were trained on the four years of ERA5 data, except one month, May 2015, which was used for testing. Due to limited space, here, we discuss results only from the global nonlocal model (M3), which is the most complex among the three models considered.

\section{Results}
Predictions of the globally resolved fluxes using model M3 show a strong agreement, both in terms of the mean climatology for May 2015 (Figure \ref{fig:flux_comparison}) and intermediate snapshots (not shown). The predictions from M3 outperform predictions from both M1 and M2, demonstrating the importance of nonlocality and model complexity in learning the nonlinear evolution of atmospheric waves. The model accurately predicts both the structure and strength of the normalized fluxes over well-known stationary GW hotspots including the Rocky Mountains, the Andes, and the East Asian Mountains. Even in the tropics, where most GWs are generated by moist convective activity, the predicted mean climatology agrees reasonably well with the normalized fluxes from WINDSET (ERA5).

The prediction skill in the tropics is relatively weaker than in the midlatitudes, as quantified by the R$^2$ metric. For the zonal flux, M3 achieves an R$^2$ $\approx$ 0.6 in the midlatitudes in both hemispheres. This value is down to 0.3-0.4 on average in the tropics. Moreover, the corresponding R$^2$ values are generally weaker for the meridional flux.

The prediction skill is quite poor in the stratosphere, where even negative R$^2$ values are obtained in the tropics. This is due to an exponential decrease in the background density and strong shear in the stratosphere, leading to data imbalance and reduced predictive skill. Efforts to enhance the prediction skill in the stratosphere are currently underway.

These results highlight (a) the challenges involved in simulating small-scale nonlocal wave evolution in the atmosphere, and (b) that simulation of non-stationary GWs can be more challenging than stationary or quasi-stationary GWs generated over orography which may have longer wavelengths.

This is work in progress and the next steps include transfer learning-focused experiments to combine the fluxes from WINDSET with GW fluxes obtained from global 1 km climate models that resolve the whole mesoscale spectrum.

\section*{Broader Impact}
Model parameterizations present as a major source of uncertainty in current climate models. Success with nonlocal ML simulation of GWs can be extended to develop ML simulators for a broad range of unresolved mesoscale and submesoscale processes in coarse-climate models, potentially reducing model uncertainty. 

These ML models can also be used as downstream plugins for weather and climate foundation models for quick and inexpensive weather forecasting and climate prediction, empowering climate model use by the wider community and for educational purposes. Efforts in this direction are underway.


\bibliography{example_paper}
\bibliographystyle{icml2024}

  
\section{Appendix}
\appendix


\begin{figure*}[ht]
    \centering
    \includegraphics[width=0.85\textwidth,keepaspectratio]{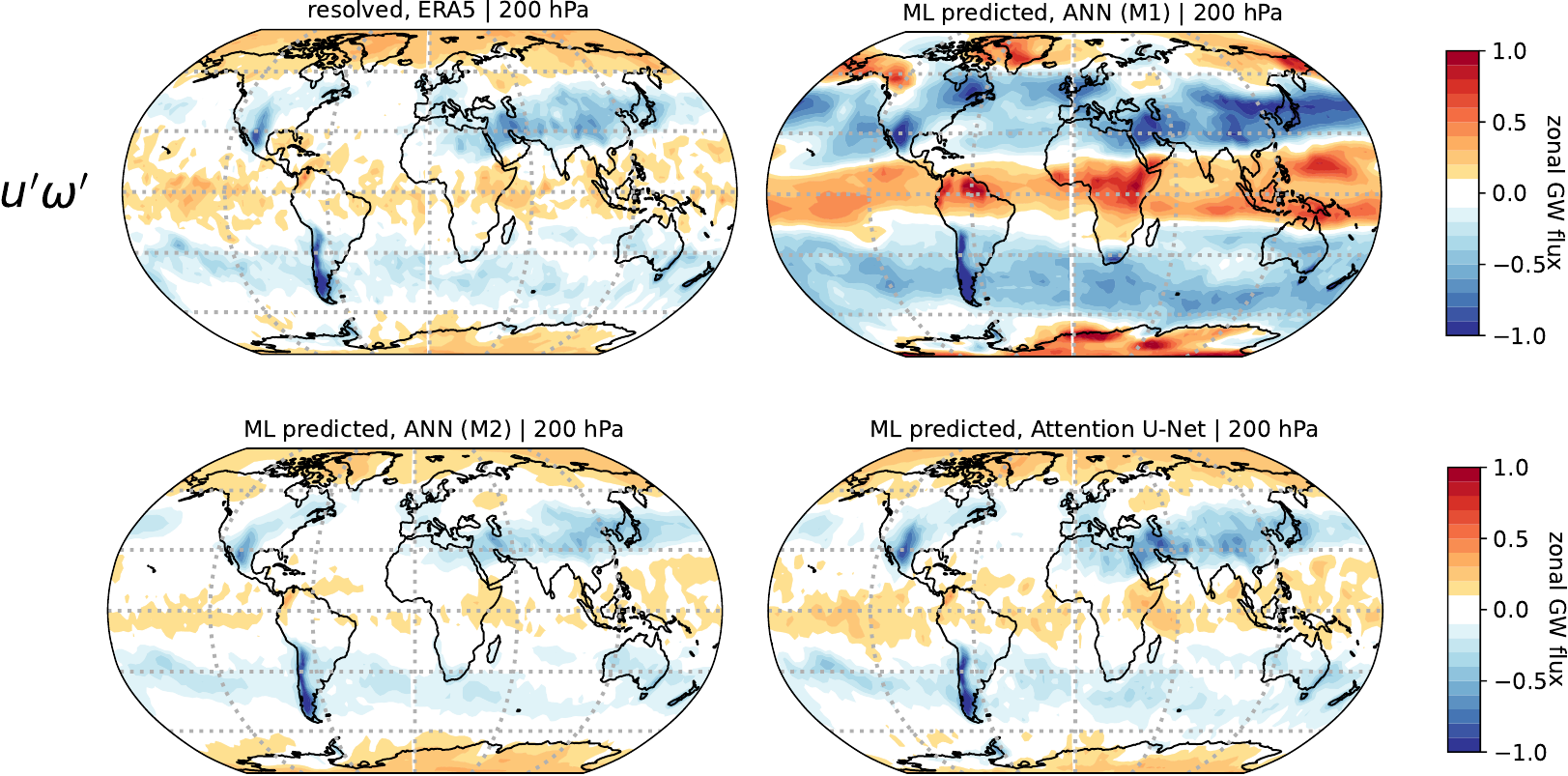}
    \caption{Mean predicted fluxes compared with the (top left) true ERA5 flux from the (top right) M1: single-column ANN, (bottom left) M2: 3x3 nonlocal columns ANN, and (bottom right) M3: globally nonlocal Attnetion U-Net CNN, for May 2015 at 200 hPa height. The figure compares the true mean and the predicted mean vertical flux of zonal momentum ($u'\omega'$) for the 3 models trained for the same number of epochs. The 1x1 and 3x3 ANNs had identical hyperparameters and the 3x3 input was processed and propagated into a single 1x1 column input by applying a 3x3 2D convolution layer. Even though the 1x1 ANN roughly captures the gross structure of the fluxes, and identifies the stationary hotspots in the midlatitudes to a certain degree, the predictions have a clear strong bias. Moreover, M1 incorrectly predicts the sign of the zonal flux over most of the Northern polar region and the sign of the meridional flux over most of the Southern polar region. Introducing nonlocal leads to a drastic improvement in performance, reduced model overfitting, and produces better prediction. The globally nonlocal UNet provides the best prediction.}
    \label{fig:comp_zonal}
\end{figure*}

\begin{figure*}[ht]
    \centering
    \includegraphics[width=0.85\textwidth,keepaspectratio]{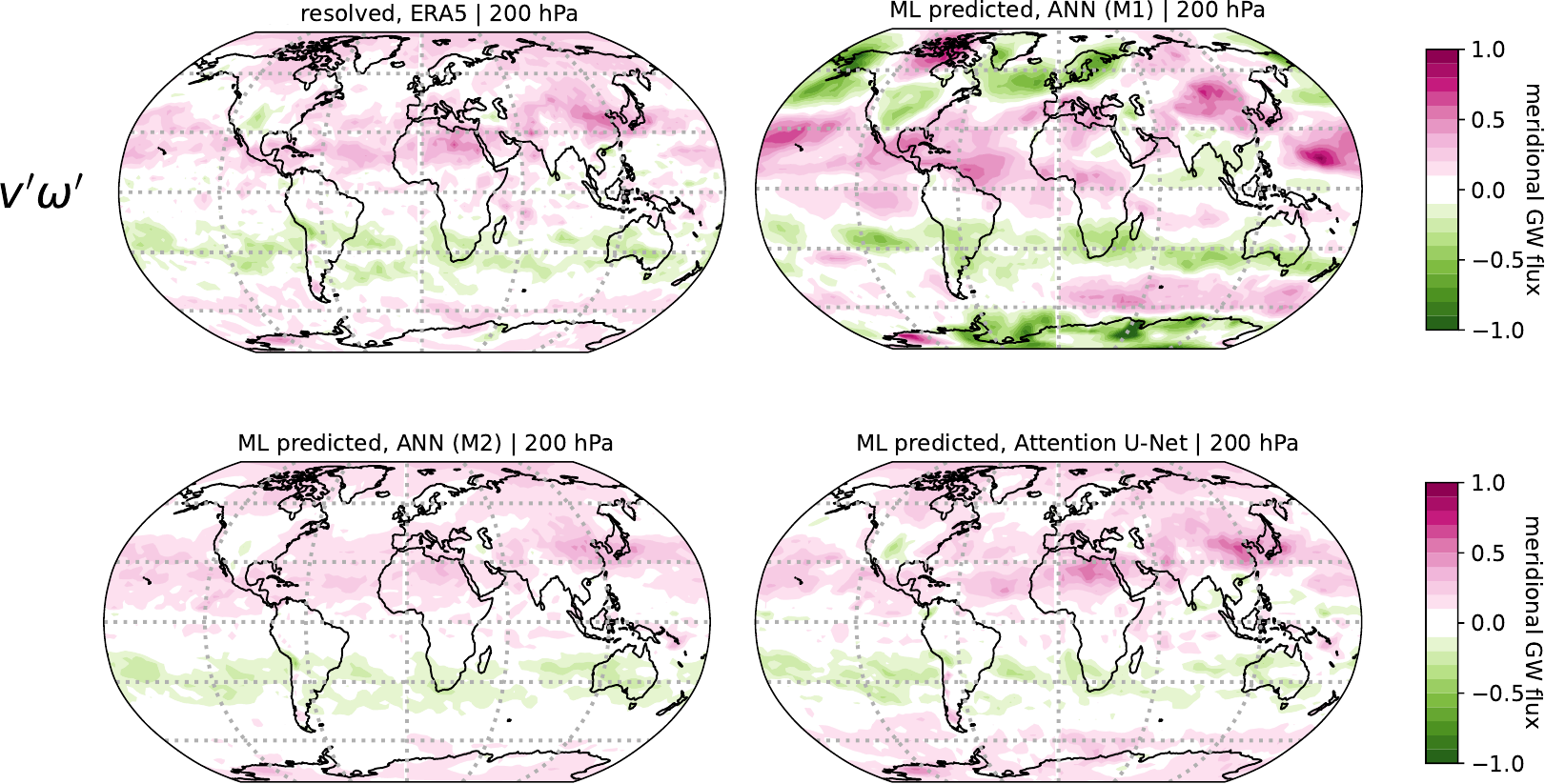}
    \caption{Same comparison as in Figure \ref{fig:comp_meridional}, but for the meridional flux of vertical momentum ($v'\omega'$) at 200 hPa.}
    \label{fig:comp_meridional}
\end{figure*}

\begin{figure*}[!ht]
    \centering
    \includegraphics[width=0.85\textwidth,keepaspectratio]{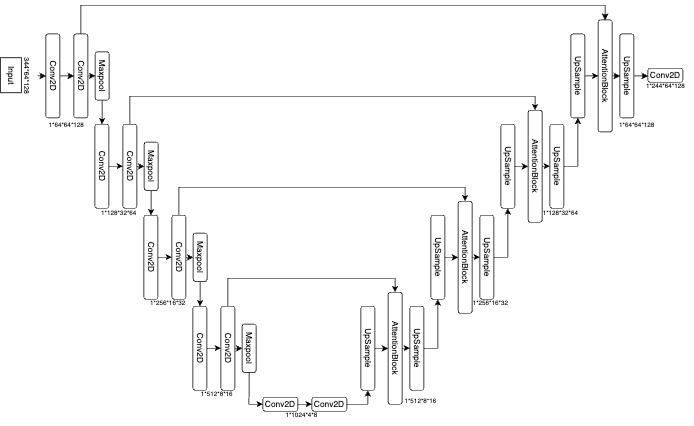}
    \caption{Schematic illustrating the architecture of the Attention-UNET used in the global nonlocal model M3.}
    \label{fig:attn_unet}
\end{figure*}

\begin{figure*}[!ht]
    \centering
    \includegraphics[width=0.8\textwidth,keepaspectratio]{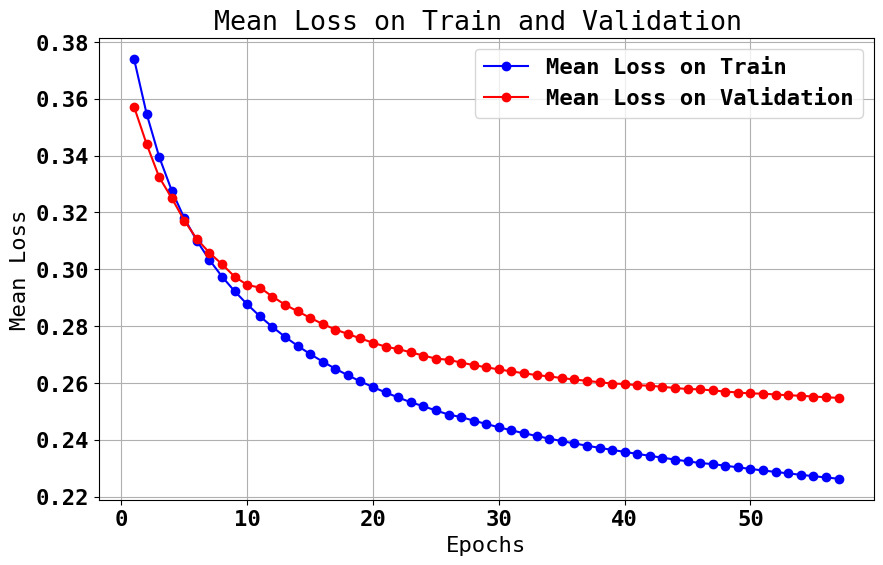}
    \caption{Training and Validation loss curve for the Attention-UNet model, M3.}
    \label{fig:attn_unet_loss}
\end{figure*}



\end{document}